\documentclass[preprint,12pt]{elsarticle}

\usepackage{amsmath,amssymb,mathtools}
\usepackage{graphicx}
\usepackage{booktabs}
\usepackage{array}
\usepackage{tikz}
\usetikzlibrary{arrows.meta,positioning,fit,calc}
\usepackage{hyperref}
\graphicspath{{figs/}}

\newcommand{\kbar}{\bar{k}}
\newcommand{\kc}{\bar{k}_{c}}
\newcommand{\Reff}{\mathcal{R}_{\mathrm{eff}}}
\newcommand{\nuT}{\nu_{T}}
\newcommand{\nuF}{\nu_{F}}
\newcommand{\nuH}{\nu_{H}}
\newcommand{\sigH}{\sigma_{H}}

\journal{Chaos, Solitons \& Fractals}

\makeatletter
\def\ps@pprintTitle{%
  \let\@oddhead\@empty\let\@evenhead\@empty
  \def\@oddfoot{\hfill\thepage\hfill}\let\@evenfoot\@oddfoot}
\makeatother

\begin{document}

\begin{frontmatter}

\title{Oscillations in two-stage contagion on networks: contagion structure, turnover and connectivity}

\author[upm]{Francisco J.\ Mu\~noz\corref{cor1}}
\ead{fcomunoz@alumnos.upm.es}
\author[upm]{Juan Carlos Nu\~no}
\author[fi,dinfo]{Luca Meacci}
\author[fi]{Mario Primicerio\textsuperscript{$\dagger$}}
\nonumnote{$^{\dagger}$Deceased.}

\cortext[cor1]{Corresponding author}
\address[upm]{Departamento de Matem\'atica Aplicada, Universidad Polit\'ecnica de Madrid, 28040 Madrid, Spain}
\address[fi]{Dipartimento di Matematica e Informatica ``Ulisse Dini'', Universit\`a degli Studi di Firenze, Viale Morgagni, 67/a, 50134, Firenze, Italy}
\address[dinfo]{Dipartimento di Ingegneria dell'Informazione (DINFO), Universit\`a degli Studi di Firenze, Via di S.~Marta, 3, 50139, Firenze, Italy}

\begin{abstract}
Two-stage contagion describes processes, such as the adoption of an opinion or of a technological novelty, in which an individual must be exposed to an adopter twice before adopting. With demographic turnover these processes can generate waves of adoption and abandonment. We study these oscillations in a two-stage contagion model with turnover through a reservoir of inactive individuals and a small rate of spontaneous activation, and report four main findings. (i) In the mean-field description the oscillations are born at a supercritical Hopf bifurcation and require a predominantly two-stage contagion (first-contact adoption probability $Q\lesssim0.17$) and a slow turnover; spontaneous activation acts as a perturbation that destroys them. (ii) In finite populations demographic noise sustains quasi-cycles also outside this region, with a spectrum and variance predicted by a linear noise approximation away from the Hopf bifurcation. (iii) Reducing the connectivity of the network damps the oscillation, because naive nodes see fewer active neighbours than in a well-mixed population, and correlations measured in simulations confirm this damping mechanism; a pair approximation predicts that the sustained cycle disappears below a critical mean degree $\kc$, which decreases as $Q$ and the turnover rate decrease, and a finite-size test with up to $3.2\times10^5$ nodes confirms this threshold in simulations. (iv) At low connectivity the amplitude no longer depends on $Q$, so that sparsity erases the signature of the two-stage mechanism, whereas a faster turnover reduces the relative amplitude at every connectivity. The irregular dynamics observed in simulations are noise-driven quasi-cycles, not chaos.
\end{abstract}


\end{frontmatter}

\section{Introduction}
\label{sec:intro}

In many social processes a single contact with an adopter is not enough to make an individual adopt: the first exposure makes the individual aware of the novelty or of the opinion, and a further exposure is needed to convert awareness into adoption. This idea is behind threshold and complex-contagion models of collective behaviour \cite{granovetter1978,dodds2004,centola2007} and is the basis of the \emph{two-stage contagion} models introduced and analysed by Katriel \cite{katriel2019}. In these models naive individuals ($S_0$) who meet an adopter ($A$) become either adopters or merely informed ($S_1$), and informed individuals become adopters on a further contact. Katriel showed that, when demographic turnover is included, two-stage contagion produces bistability, critical transitions, endogenous oscillations and excitability, none of which occur in the classical one-stage epidemic models \cite{katriel2019}; a numerical bifurcation analysis of a related model has also been reported \cite{heidecke2020}.

Endogenous oscillations are perhaps the most striking of these phenomena from the point of view of applications, since they correspond to recurrent waves of adoption and abandonment of a fashion, an opinion or a technology without any periodic external forcing. Three questions are natural. First, which features of the contagion and of the population renewal produce the oscillations, and how robust are they in parameter space, in particular to a small exogenous (spontaneous) adoption, which is present in any real setting through media or advertising? Second, what happens in finite populations, where demographic noise can either destroy a deterministic cycle by extinction or, conversely, sustain oscillations where the deterministic dynamics is damped \cite{mckane2005,alonso2007}? Third, how do the oscillations depend on the structure of the contact network, in particular on its connectivity?

Spatial effects on two-stage contagion have been addressed with cellular automata \cite{meacci2025}, and for one-stage epidemics on Erd\H{o}s--R\'enyi networks the validity of the mean-field description has been examined as a function of the average connectivity \cite{munoz2025}. Here we address the three questions above for a two-stage model with turnover through a reservoir of inactive individuals and spontaneous activation, on networks of different connectivity and topology. We combine (i) a bifurcation analysis of the mean-field equations, (ii) a linear noise approximation of the stochastic dynamics, (iii) a pair approximation that incorporates the mean degree and (iv) stochastic simulations on Erd\H{o}s--R\'enyi networks and, in the Supplementary Material, on random-regular, Barab\'asi--Albert and Watts--Strogatz networks.

The paper is organised as follows. Section~\ref{sec:model} defines the model. Section~\ref{sec:mf} analyses the oscillations of the mean-field equations. Section~\ref{sec:noise} studies noise-driven oscillations on the complete graph, i.e.\ in finite, well-mixed populations. Section~\ref{sec:conn} studies the effect of the network connectivity with the pair approximation and simulations, including the neighbour correlations that produce the damping, a finite-size test of the critical connectivity, the interplay of the connectivity with $Q$ and $\nu$ and a summary of the effect of the topology. Section~\ref{sec:other} briefly reports bistability and the absence of chaos, and Section~\ref{sec:disc} discusses the results and concludes.

\section{The model}
\label{sec:model}

\subsection{Rules}

\begin{figure}[!tbp]
\centering
\begin{tikzpicture}[>=Stealth, every node/.style={font=\small}, st/.style={circle,draw,minimum size=9mm,inner sep=0pt}]
\node[st] (S0) {$S_0$};
\node[st,right=3.0cm of S0] (A) {$A$};
\node[st,below=2.0cm of $(S0)!0.5!(A)$] (S1) {$S_1$};
\node[st,right=3.0cm of A] (R) {$R$};
\node[draw,rounded corners,dashed,fit=(S0)(S1)(A)(R),inner sep=7mm,label={[font=\footnotesize,yshift=1mm]above:non-inactive population (fraction $w$)}] (box) {};
\node[st,left=2.6cm of box.west] (I) {$I$};
\draw[->] (S0) -- node[above]{$Q\varphi$} (A);
\draw[->] (S0) -- node[below left,inner sep=1pt]{$(1-Q)\varphi$} (S1);
\draw[->] (S1) -- node[below right,inner sep=1pt]{$\varphi$} (A);
\draw[->] (A) -- node[above]{$\gamma$} (R);
\draw[->,dotted] (S0) to[bend left=30] node[above]{$\sigma$} (A);
\draw[->] (I.north) to[out=70,in=180] node[above left,font=\footnotesize,inner sep=1pt]{$\nu w/(1-w)$} (S0.west);
\draw[->] (box.west |- I.east) -- node[below,font=\footnotesize]{$\nu$} (I.east);
\end{tikzpicture}
\caption{Scheme of the model. $\varphi$ is the fraction of active neighbours. Solid arrows with $\varphi$: transitions triggered by contact with active neighbours (the two stages of the contagion); $\gamma$: end of adoption; dotted arrow: spontaneous activation; $\nu$: turnover, i.e.\ each non-inactive node, whatever its state, becomes inactive at rate $\nu=m/(N-n_I)$, and inactive nodes re-enter as naive ($S_0$) at the rate that keeps $n_I$ constant.}
\label{fig:scheme}
\end{figure}
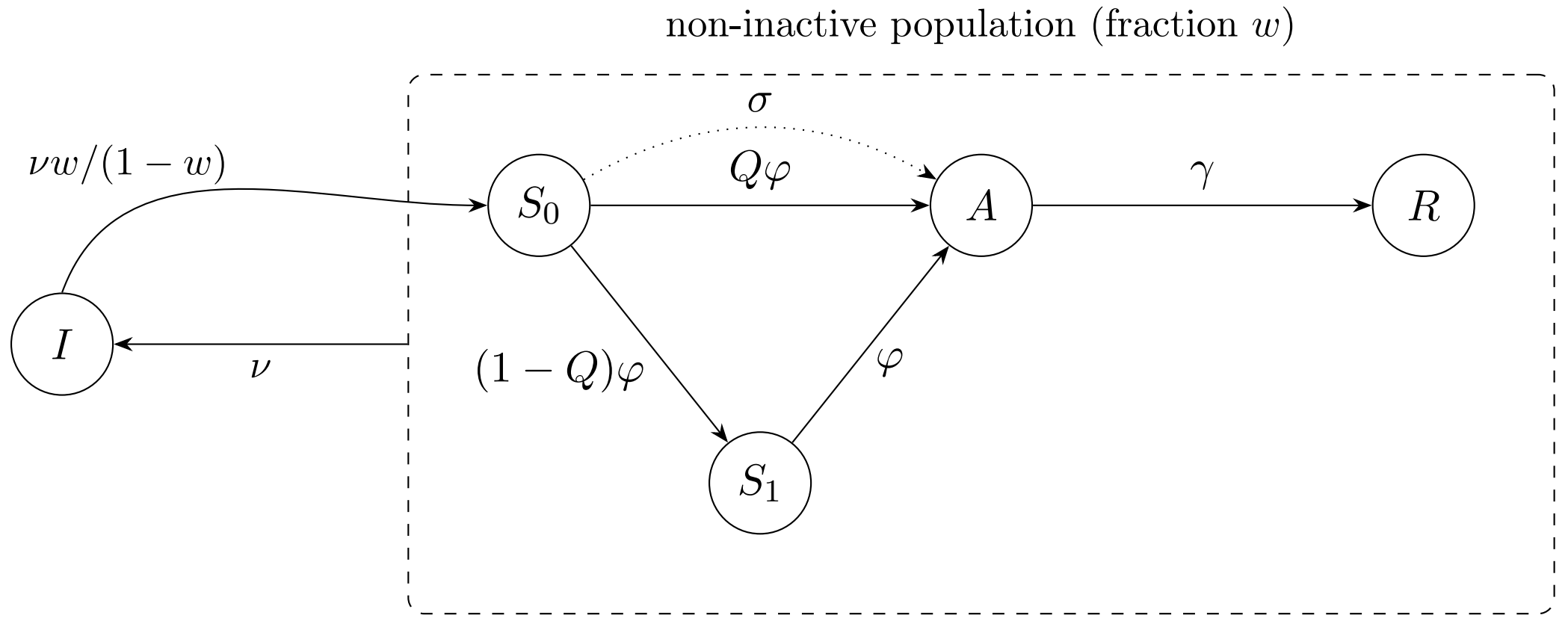

A network of $N$ nodes is given. Each node is in one of five states: naive $S_0$, informed $S_1$, active (adopter) $A$, refractory (former adopter) $R$, and inactive $I$. Inactive nodes do not take part in the contagion and act as a reservoir that feeds the turnover. Time is discrete and, at each step, all nodes are updated synchronously according to three rules (Fig.~\ref{fig:scheme}).

\begin{enumerate}
\item \textbf{Contagion.} Let $\varphi_i=m_i/k_i$ be the fraction of neighbours of node $i$ that are in state $A$ ($m_i$ active neighbours among $k_i$; $\varphi_i=0$ if $k_i=0$). A node in $S_0$ becomes $A$ with probability $Q\varphi_i$ and $S_1$ with probability $(1-Q)\varphi_i$; a node in $S_1$ becomes $A$ with probability $\varphi_i$; a node in $A$ becomes $R$ with probability $\gamma$. Nodes in $R$ and $I$ do not change.
\item \textbf{Turnover.} $m=\lceil\mu\,(N-n_I)\rceil$ nodes chosen at random among the non-inactive nodes become $I$, and $m$ nodes chosen at random among the inactive ones become $S_0$. The number $n_I$ of inactive nodes is therefore constant.
\item \textbf{Spontaneous activation.} Each node in $S_0$ becomes $A$ with probability $\sigma$.
\end{enumerate}

Because the transition probabilities depend on the \emph{fraction} of active neighbours, a node with few neighbours is as likely to be influenced as a node with many, and the degree enters only through the discreteness and the correlations of the neighbourhood. This is the feature that makes the mean-field description independent of the connectivity (Section~\ref{sec:mf}) and that the pair approximation of Section~\ref{sec:conn} corrects. The parameter $Q$ measures how much of the contagion is of the one-stage type: $Q=1$ is a one-stage (SIR-like) contagion with turnover, and $Q\to0$ a purely two-stage one.

\subsection{Parameters, networks and observables}

Unless stated otherwise we use $N=5000$, $n_I=0.75N$ (so $w=1-n_I/N=0.25$), $Q=0.15$, $\gamma=0.03125$, $\mu=0.00125$ and one initial active node, the others being $S_0$ or $I$ at random. With these values $m=2$ nodes are exchanged per step, i.e.\ a turnover rate $\nu=m/(N-n_I)=0.0016$ per non-inactive node and per step. Because of the ceiling, the reference values give $\nu=0.0016$ rather than $\mu=0.00125$; $\nu$ is the parameter of all the equations below, and $\mu$ plays no further role. When $N$ is varied, $m=\nu(N-n_I)=\nu wN$ is kept proportional to $N$ so that $\nu$ is fixed (with $m=\lceil\mu(N-n_I)\rceil$ the rate would change with $N$, e.g.\ $m=7$ and $\nu=0.0014$ for $N=20000$), and when $\nu$ is varied at $N=5000$ we use $m=1,2,4,6$ for $\nu=0.0008,0.0016,0.0032,0.0048$. We call these values the \emph{reference parameters}, and vary the first-contact adoption probability $Q$, the turnover rate $\nu$, the spontaneous activation $\sigma\in[10^{-5},10^{-3}]$, the network size, the mean degree $\kbar$ and the topology.

The simulations of the main text use Erd\H{o}s--R\'enyi (ER) graphs $G(N,p)$, with mean degree $\kbar=p(N-1)$, and the complete graph ($p=1$). Random regular, Barab\'asi--Albert \cite{barabasi1999} and Watts--Strogatz \cite{watts1998} networks are considered in the Supplementary Material.

For each run we record the number $A(t)$ of active nodes and, after a transient, compute its mean and standard deviation (the latter is our measure of the oscillation amplitude), the fraction of time with $A=0$, and the power spectral density $S(\omega)=\sum_\tau C(\tau)e^{-i\omega\tau}$ of $A(t)$ ($C$ is the autocovariance), estimated with Welch's method \cite{welch1967} (segments of $2^{14}$ steps). From the spectrum we obtain the dominant period, either as $2\pi/\omega^*$, with $\omega^*$ the position of the spectral peak, or, when the peak is broad, as $2\pi/\bar\omega$, with $\bar\omega$ the centroid of the spectrum in the band of periods $250$--$3000$ steps. We call the ratio of the standard deviation to the mean the \emph{relative amplitude}.

\section{Mean-field oscillations}
\label{sec:mf}

\subsection{Equations, thresholds and equilibria}

Let $s_0,s_1,a,r$ denote the fractions of nodes in $S_0,S_1,A,R$, with $s_0+s_1+a+r=w$. Assuming that every non-active node sees on average a fraction $a$ of active neighbours (complete graph or very dense network) and passing to continuous time ($\Delta t=1$, small probabilities per step), the expected dynamics is
\begin{align}
\dot s_0 &= \nu w-(a+\sigma+\nu)\,s_0, \label{eq:s0}\\
\dot s_1 &= (1-Q)\,a\,s_0-(a+\nu)\,s_1, \label{eq:s1}\\
\dot a &= Q\,a\,s_0+\sigma\,s_0+a\,s_1-(\gamma+\nu)\,a, \label{eq:a}\\
\dot r &= \gamma\,a-\nu\,r. \label{eq:r}
\end{align}
Since $r$ does not feed back, the dynamics is three-dimensional in $(s_0,s_1,a)$. The system does not depend on the connectivity of the network. For $\sigma=0$, rescaling the populations by $w$ leaves a system with the structure of the temporary-adoption two-stage model of Ref.~\cite{katriel2019}, with contact parameter $w/(\gamma+\nu)$ and first-contact adoption probability $Q$.

For $\sigma=0$ the state $a=0$ is an equilibrium, unstable when
\begin{equation}
\Reff=\frac{Qw}{\gamma+\nu}>1 \quad\Longleftrightarrow\quad \nu<\nuT\equiv Qw-\gamma ,
\label{eq:Reff}
\end{equation}
($\Reff=1.14$ for the reference parameters). The endemic equilibria satisfy $s_0^*=\nu w/(a^*+\nu)$, $s_1^*=(1-Q)a^*s_0^*/(a^*+\nu)$ and
\begin{equation}
(\gamma+\nu)\,(a^*+\nu)^2=\nu w\,(a^*+Q\nu),
\label{eq:quad}
\end{equation}
and two positive roots coexist with the stable extinction state for $\nuT<\nu<\nuF\equiv w/[4(1-Q)]-\gamma$. This bistability is summarised in Section~\ref{sec:other} and analysed in Section~S3 of the Supplementary Material. For $\sigma>0$ the extinction state disappears; $a^*$ is the positive root of $(Qa^*+\sigma)s_0^*+a^*s_1^*=(\gamma+\nu)a^*$ with $s_0^*=\nu w/(a^*+\sigma+\nu)$, obtained numerically. For the reference parameters and $\sigma=10^{-4}$, $Na^*\approx45$ nodes and the eigenvalues of the Jacobian are $-0.0215$ and $-1.5\times10^{-4}\pm0.0113\,i$: a weakly damped focus with period $2\pi/0.0113\approx555$ steps and decay time $\approx6500$ steps.

\subsection{The oscillatory region: contagion structure and turnover}

\begin{figure}[!tbp]
\centering
\includegraphics[width=\textwidth]{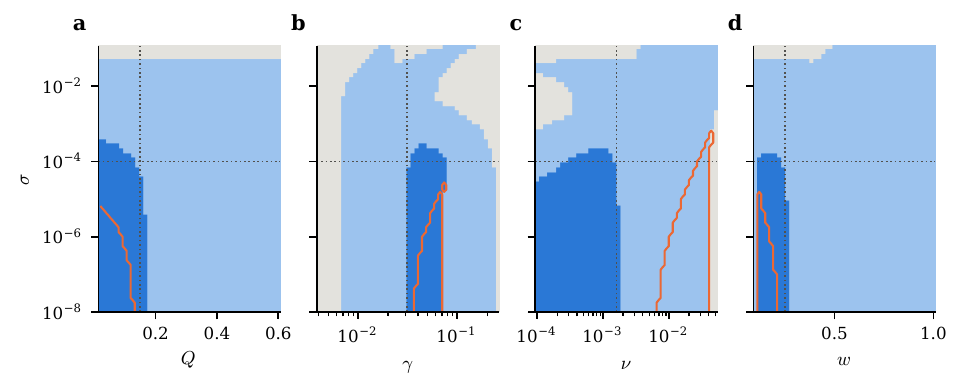}
\caption{Mean-field stability of the positive equilibrium in the planes (a) $(Q,\sigma)$, (b) $(\gamma,\sigma)$, (c) $(\nu,\sigma)$ and (d) $(w,\sigma)$; the remaining parameters are at their reference values (dotted lines). Dark blue: unstable focus with a stable limit cycle (sustained oscillations); light blue: stable focus (damped oscillations); grey: stable node. The orange contour encloses the region with more than one positive equilibrium.}
\label{fig:mfmaps}
\end{figure}

Fig.~\ref{fig:mfmaps} shows the stability of the equilibrium in the planes $(Q,\sigma)$, $(\gamma,\sigma)$, $(\nu,\sigma)$ and $(w,\sigma)$, the other parameters being kept at their reference values. The vertical direction of each map ($\sigma\to0$) shows which features of the contagion and of the renewal of the population produce the oscillation; the horizontal direction shows how much spontaneous activation it tolerates. Sustained oscillations (dark region) occupy a bounded region of each plane:
\begin{itemize}
\item \textbf{Two-stage contagion is needed.} For $\sigma\to0$ the equilibrium is unstable only for $Q\lesssim0.165$; for larger $Q$ the dynamics is a damped focus, as in one-stage models with turnover. The Hopf threshold $\sigH$ increases as $Q$ decreases, up to $\sigH\approx3.7\times10^{-4}$ at $Q=0.02$. Thus the more the contagion relies on the second stage, the more robust the oscillation is to spontaneous activation. The informed class introduces a delay: after a wave, $S_1$ nodes accumulate and convert into adopters only when activity reappears, which produces the overshoot.
\item \textbf{Recovery must be neither too slow nor too fast.} Oscillations require $0.030\lesssim\gamma\lesssim0.07$; for smaller $\gamma$ the system is far above threshold and strongly damped, and for larger $\gamma$ the endemic equilibrium disappears at the fold that bounds the bistable region (orange contour).
\item \textbf{Slow turnover.} Oscillations require $\nu\lesssim\nuH\approx0.0017$ (the $\sigma=0$ Hopf point), with the largest $\sigH\approx2\times10^{-4}$ at $\nu\approx9\times10^{-4}$. The turnover is the slow process that refills the naive population and sets the period of the relaxation oscillations.
\item \textbf{A large inactive reservoir.} Oscillations require $w\lesssim0.26$, i.e.\ more than $74\%$ of inactive nodes.
\end{itemize}
The reference parameters lie close to the boundary in $Q$, $\gamma$, $\nu$ and $w$, which explains why, for these parameters, a spontaneous activation as small as $4\times10^{-5}$ is enough to destroy the cycle. Over the whole explored region the largest Hopf threshold is $\sigH\approx3.7\times10^{-4}$; for stronger spontaneous activation the dynamics is a damped focus and, for $\sigma\gtrsim0.05$, a node. Damped oscillations are not irrelevant, however: in finite populations they are excited by the noise.

\subsection{Spontaneous activation as a perturbation}

\begin{figure}[!tbp]
\centering
\includegraphics[width=0.9\textwidth]{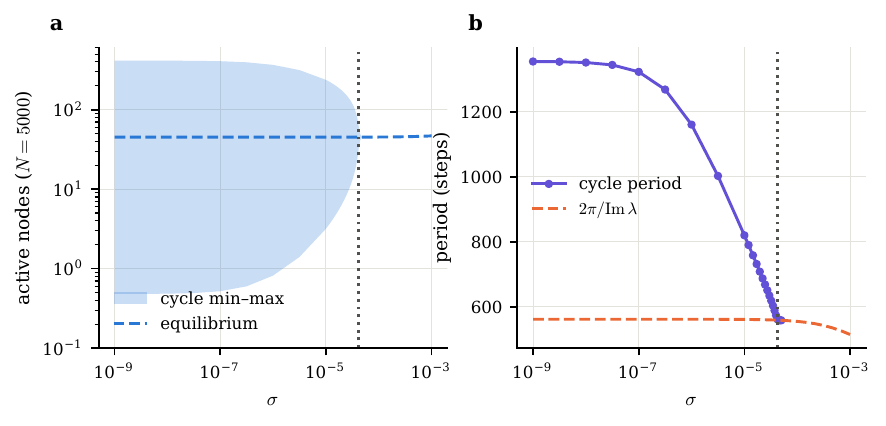}
\caption{Mean-field limit cycle versus the spontaneous activation $\sigma$ (reference parameters; fractions converted into nodes with $N=5000$). (a) Equilibrium (dashed) and range of the stable limit cycle (shaded); the dotted line marks $\sigH=4.2\times10^{-5}$. (b) Period of the cycle (symbols) and linear period $2\pi/\mathrm{Im}\,\lambda$ of the focus (dashed).}
\label{fig:cycle}
\end{figure}

Spontaneous activation does not create the oscillation but competes with it: it maintains a minimum flow of new adopters that fills the troughs of the cycle and stabilises the equilibrium. The leading complex pair crosses the imaginary axis at $\sigH=4.20\times10^{-5}$ (reference parameters): for $\sigma<\sigH$ the equilibrium is an unstable focus surrounded by a stable limit cycle. The bifurcation is supercritical: integrations started near the equilibrium and from one active node converge to the same attractor (no hysteresis), and close to $\sigH$ the squared amplitude of the cycle grows linearly with $\sigH-\sigma$ (Fig.~\ref{fig:cycle}a). The period of the cycle is $\approx560$ steps at the bifurcation, close to the linear value $2\pi/\mathrm{Im}\,\lambda$, and increases as $\sigma$ decreases, to $820$ steps at $\sigma=10^{-5}$ and $\approx1350$ steps for $\sigma\to0$ (Fig.~\ref{fig:cycle}b). In the same limit the minimum of the cycle falls far below one node: the oscillation becomes a relaxation oscillation in which a burst of adoption depletes the naive population, which is then slowly replenished by the turnover until a new burst can start. A finite population cannot follow such deep troughs (Section~\ref{sec:noise}). Note that $\sigH$ refers to the continuous-time equations; the linearisation of the synchronous update, $\mathbf{x}_{t+1}\simeq\mathbf{x}_t+J(\mathbf{x}_t-\mathbf{x}^*)$, has eigenvalues $1+\lambda$, whose modulus exceeds one for $\sigma<\sigH^{(d)}\approx6.6\times10^{-5}$. This shift matters only very close to the bifurcation, but it is included in the noise analysis of Section~\ref{sec:noise}.

\section{Noise-driven oscillations on the complete graph}
\label{sec:noise}

\subsection{Linear noise approximation}

In a finite population the number of active nodes fluctuates around the deterministic trajectory. Above the Hopf bifurcation, when the equilibrium is a weakly damped focus, these fluctuations are filtered by the linear dynamics and produce \emph{quasi-cycles}: irregular oscillations whose spectrum has a peak close to the frequency of the focus \cite{mckane2005,alonso2007}. We describe them with a linear noise approximation (LNA) of the synchronous update. Writing the state as $\mathbf{x}=(s_0,s_1,a)=\mathbf{x}^*+N^{-1/2}\boldsymbol\xi$, the fluctuations obey, to leading order in $N^{-1/2}$, the linear map
\begin{equation}
\boldsymbol\xi_{t+1}=B\,\boldsymbol\xi_t+\boldsymbol\eta_t,\qquad B=\mathbb{I}+J,
\label{eq:lna}
\end{equation}
where $J$ is the Jacobian of Eqs.~(\ref{eq:s0}--\ref{eq:a}) at the equilibrium and $\boldsymbol\eta_t$ is a white noise with covariance $D$ given by the transitions of the model (\ref{app:lna}). The stationary covariance $\Sigma$ of $\boldsymbol\xi$ solves the discrete Lyapunov equation $\Sigma=B\Sigma B^{\mathsf T}+D$, and the spectrum of the number of active nodes is
\begin{equation}
S_A(\omega)=N\left[G(\omega)\,D\,G(\omega)^{\dagger}\right]_{aa},\qquad G(\omega)=\left(e^{i\omega}\mathbb{I}-B\right)^{-1}.
\label{eq:spec}
\end{equation}
The LNA predicts that the variance of $A$ grows as $N$ (the relative amplitude decreases as $N^{-1/2}$) and that the spectral peak is at $\omega^*\simeq\arg(1+\lambda)\simeq\mathrm{Im}\,\lambda$ with a width $\simeq2(1-|1+\lambda|)$, so that the coherence of the quasi-cycle, measured by the quality factor $\omega^*/\Delta\omega$, is independent of $N$: $63$ for $\sigma=10^{-4}$, $9.6$ for $\sigma=3\times10^{-4}$ and $2.8$ for $\sigma=10^{-3}$.

\subsection{Comparison with simulations}

\begin{figure}[!tbp]
\centering
\includegraphics[width=\textwidth]{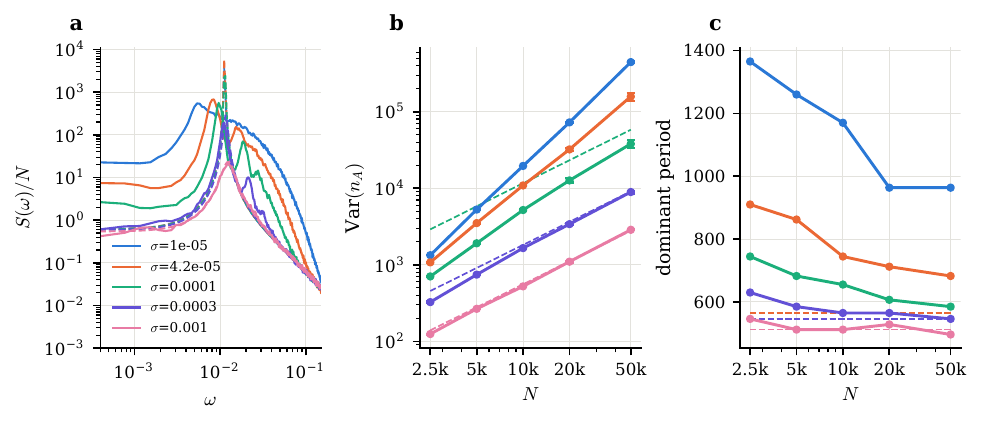}
\caption{Noise-driven oscillations on the complete graph. (a) Power spectral density of $A(t)$ divided by $N$ for $N=10000$ and five values of $\sigma$ (solid: simulations; dashed: LNA, Eq.~\ref{eq:spec}; the LNA is not defined below $\sigH^{(d)}$, i.e.\ for the two smallest values of $\sigma$). (b) Variance of $A$ versus $N$ (symbols: simulations, with standard errors; dashed: LNA). (c) Dominant period versus $N$ (symbols: simulations; dashed: LNA peak). Colours as in (a).}
\label{fig:lna}
\end{figure}

We simulated the complete graph for $N=2500$--$50000$ and $\sigma=10^{-5},4.2\times10^{-5},10^{-4},3\times10^{-4},10^{-3}$ ($1.2\times10^5$ steps, the first $2\times10^4$ discarded; 20 runs per point for $N\le10^4$ and 5 for $N=2\times10^4$ and $5\times10^4$). Fig.~\ref{fig:lna} compares the results with the LNA.

\begin{itemize}
\item \textbf{Far from the Hopf bifurcation} ($\sigma\ge3\times10^{-4}$) the LNA is quantitative: the simulated spectra coincide with Eq.~(\ref{eq:spec}) (Fig.~\ref{fig:lna}a), and the standard deviation of $A$ is within $5\%$ of the prediction for $\sigma=10^{-3}$ and $N\ge5000$, and for $\sigma=3\times10^{-4}$ and $N\ge10^4$ (e.g.\ $16.3\pm0.1$ against $16.7$ nodes for $N=5000$, $\sigma=10^{-3}$, and $94.4\pm2.2$ against $95.4$ for $N=50000$, $\sigma=3\times10^{-4}$); the agreement improves with $N$. For $N\ge5000$ the mean activity agrees with the mean-field equilibrium within $2\%$.
\item \textbf{Close to the Hopf bifurcation} ($\sigma=10^{-4}$, i.e.\ $2.4\,\sigH$ and $1.5\,\sigH^{(d)}$) the quasi-cycles are large: the standard deviation is $44$ nodes for a mean of $42$ at $N=5000$. The fluctuations are then not small and the LNA overestimates the variance (standard deviation $76$ against $44$ nodes at $N=5000$), although the ratio approaches one as $N$ grows ($241$ against $193\pm12$ at $N=50000$). The dominant period is longer than the linear one ($683$ steps against $555$ at $N=5000$) and decreases towards it with $N$ ($585$ steps at $N=50000$). This is consistent with Fig.~\ref{fig:cycle}b: large excursions explore the nonlinear part of the flow, where the period is longer.
\item \textbf{Below the Hopf bifurcation} ($\sigma=10^{-5}$) the deterministic cycle is reached and the variance grows as $N^2$ rather than $N$ (Fig.~\ref{fig:lna}b). The deep troughs of the cycle, however, make the activity die out frequently, to be restarted by spontaneous activation: the fraction of time with $A=0$ is $0.58$ at $N=2500$, $0.38$ at $N=5000$ and $0.001$ at $N=50000$. These extinction--reignition events lengthen the period ($1370$ steps at $N=2500$ and $\approx960$ at $N=50000$, against $820$ for the deterministic cycle) and reduce the mean activity below the mean-field equilibrium.
\end{itemize}
In summary, finite populations oscillate on both sides of the Hopf bifurcation; the bifurcation separates a regime in which the amplitude is set by the deterministic cycle (standard deviation $\propto N$) from a regime in which it is set by the noise ($\propto N^{1/2}$), and the coherence of the noise-driven oscillation is controlled by the ratio of the damping to the frequency of the focus.

\section{Effect of the network connectivity}
\label{sec:conn}

\subsection{Pair approximation}

The only place where the network enters the expected dynamics is the contagion probability. If $\varphi_X$ denotes the average, over the nodes in state $X$, of the fraction of active neighbours, the expected dynamics reads exactly
\begin{align}
\dot s_0 &= \nu w-(\varphi_{S_0}+\sigma+\nu)\,s_0, \\
\dot s_1 &= (1-Q)\,\varphi_{S_0}\,s_0-(\varphi_{S_1}+\nu)\,s_1, \\
\dot a &= Q\,\varphi_{S_0}\,s_0+\sigma\,s_0+\varphi_{S_1}\,s_1-(\gamma+\nu)\,a,
\end{align}
and the mean-field system~(\ref{eq:s0}--\ref{eq:a}) corresponds to the closure $\varphi_X=a$. In a sparse network the states of neighbouring nodes are correlated, so $\varphi_{S_0},\varphi_{S_1}\neq a$: an active node has recently been activated by, or has activated, its neighbours, which are therefore less often naive. We determine $\varphi_X$ with a pair approximation (PA): let $\pi_{XY}$ be the fraction of edge ends joining a node in state $X$ to a node in state $Y$, $n_X=\sum_Y\pi_{XY}$, $q_{Y|X}=\pi_{XY}/n_X$ and $\varphi_X=q_{A|X}$. Assuming that the $\kbar$ neighbours of a node in state $X$ are independent with state probabilities $q_{Y|X}$ (the graph is treated as regular with degree $\kbar$), the equations for $\pi_{XY}$ follow from the transitions of the model (\ref{app:pa}). The only network parameter is $\kbar$, and the limit $\kbar\to\infty$ is the mean-field system. We computed the equilibria of the PA by Newton's method, continued from the mean-field solution at large $\kbar$, and their stability from the eigenvalues of the numerically differentiated Jacobian.

\subsection{Connectivity damps the oscillation and moves the Hopf bifurcation}
\label{sec:pahopf}

\begin{figure}[!tbp]
\centering
\includegraphics[width=0.9\textwidth]{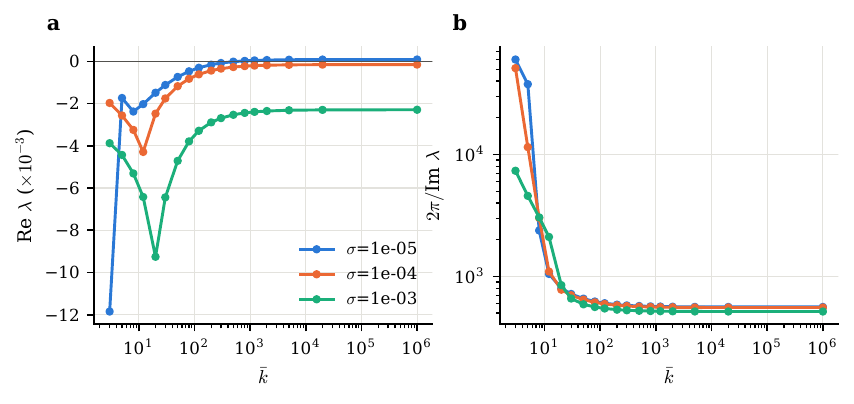}
\caption{Leading oscillatory eigenvalue of the pair-approximation equilibrium versus the mean degree $\kbar$ (reference parameters): (a) real part, (b) period $2\pi/\mathrm{Im}\,\lambda$. The rightmost points ($\kbar=10^6$) coincide with the mean-field values.}
\label{fig:paeig}
\end{figure}

Fig.~\ref{fig:paeig} shows the leading complex eigenvalue of the PA equilibrium as a function of $\kbar$. As the connectivity decreases, $\mathrm{Re}\,\lambda$ becomes more negative: at $\sigma=10^{-4}$ it goes from $-1.5\times10^{-4}$ in the mean-field limit to $-1.2\times10^{-3}$ at $\kbar=50$ and $-4.3\times10^{-3}$ at $\kbar=12$, while the period increases from $555$ to $644$ steps at $\kbar=50$ and to $\approx1100$ steps at $\kbar=12$. For $\kbar\lesssim10$ the period grows by orders of magnitude and the mode is effectively overdamped. Correlations therefore act as an additional damping. For $\sigma=10^{-5}$ the real part changes sign at a critical mean degree $\kc$ between $535$ and $598$: below $\kc$ the PA has no limit cycle. Finite-size simulations confirm this threshold quantitatively, and near and above $\kc$ the PA cycle also gives the amplitude of the oscillation within about $10\%$ (Section~\ref{sec:size}).

\begin{figure}[!tbp]
\centering
\includegraphics[width=\textwidth]{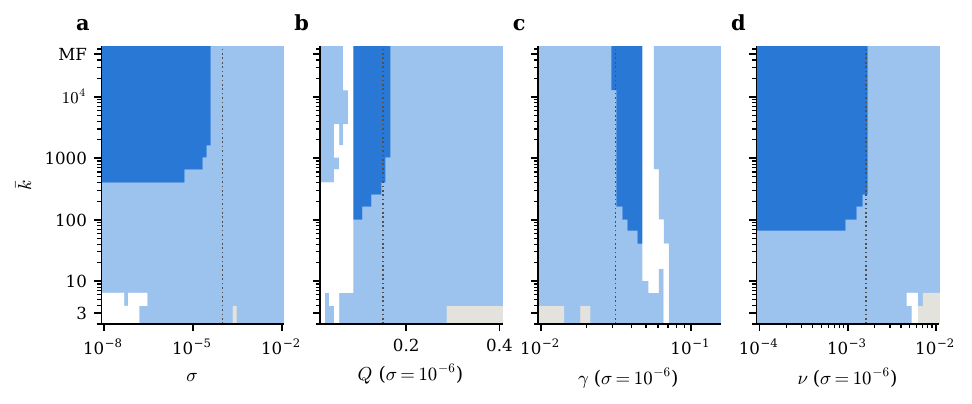}
\caption{Stability of the pair-approximation equilibrium in the planes (a) $(\sigma,\kbar)$, (b) $(Q,\kbar)$, (c) $(\gamma,\kbar)$ and (d) $(\nu,\kbar)$; in (b)--(d) $\sigma=10^{-6}$. Colours as in Fig.~\ref{fig:mfmaps}; the top row (MF) is the mean-field limit; blank cells: Newton continuation not converged. Dotted lines: reference values.}
\label{fig:pamaps}
\end{figure}

The value of $\kc$ depends strongly on the other parameters (Fig.~\ref{fig:pamaps}). For $\sigma\to0$ ($\sigma=10^{-8}$) and the reference values, $\kc$ is between $300$ and $500$, but it falls below $100$ for a slower turnover ($\nu\lesssim9\times10^{-4}$) or a faster recovery ($\gamma\approx0.042$), i.e.\ in the parameter regions where the mean-field oscillation is more robust (Fig.~\ref{fig:mfmaps}). In terms of $Q$, the oscillatory region of the PA shrinks as $\kbar$ decreases, and the more two-stage the contagion, the lower $\kc$: at $\sigma=10^{-6}$ (Fig.~\ref{fig:pamaps}b), $\kc$ is between $429$ and $479$ for $Q=0.15$, between $113$ and $126$ for $Q=0.10$ and between $65$ and $72$ for $Q=0.05$, and for $Q\gtrsim0.17$ there is no cycle at any $\kbar$, as in the mean field. The same ordering holds for the turnover (Fig.~\ref{fig:pamaps}d): $\kc$ is between $65$ and $72$ for $\nu=0.0008$, and there is no cycle for $\nu\gtrsim\nuH$. Thus the mechanism that produces the oscillation in the mean-field limit (a small $Q$ and a slow $\nu$) is also what makes it robust to sparsity. In all the explored planes the oscillatory region of the PA is contained in the mean-field one: within the PA, reducing the connectivity never creates sustained oscillations. (In the small-$Q$ and bistable regions the Newton continuation did not always converge; these points are left blank in Fig.~\ref{fig:pamaps}. At very small $\kbar$ the continuation can also land on a spurious root close to extinction; such points were recomputed starting Newton's method from a direct integration of the PA.)

\subsection{Simulations on Erd\H{o}s--R\'enyi networks}

\begin{figure}[!tbp]
\centering
\includegraphics[width=\textwidth]{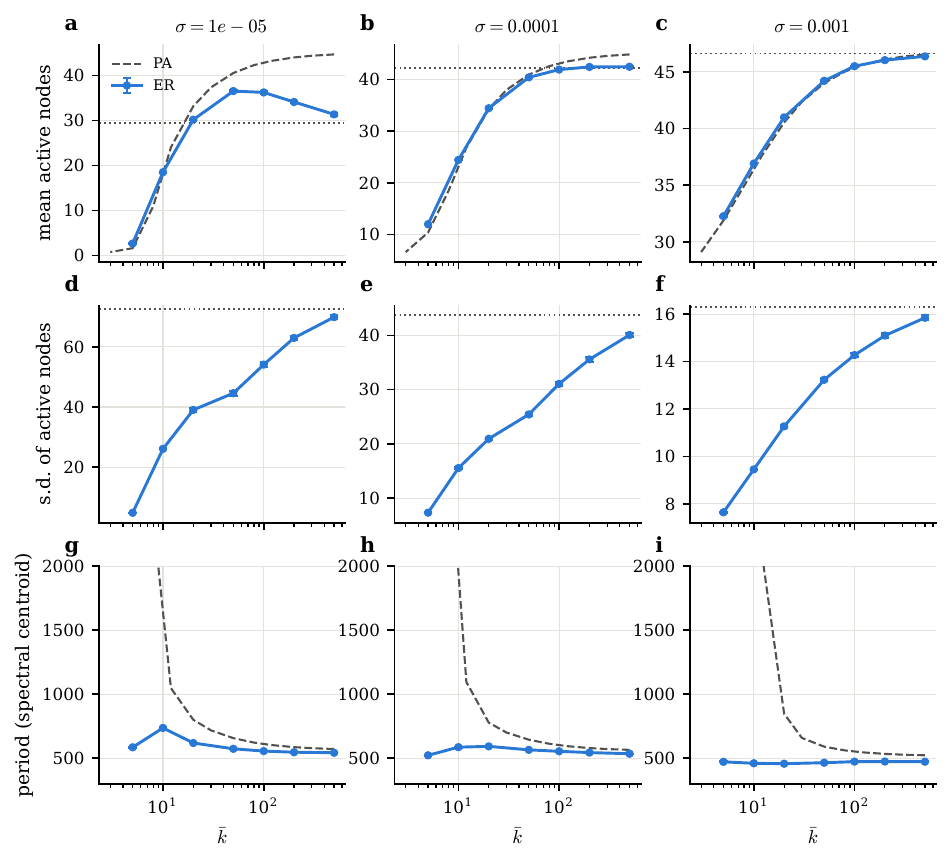}
\caption{Simulations on ER networks ($N=5000$, $10^5$ steps, 20 runs per point; error bars: standard error) versus the mean degree, for $\sigma=10^{-5}$ (left), $10^{-4}$ (centre) and $10^{-3}$ (right). (a--c) Mean number of active nodes; dashed: pair approximation. (d--f) Standard deviation of the number of active nodes. (g--i) Dominant period (centroid of the spectrum in the band of periods $250$--$3000$); dashed: PA linear period $2\pi/\mathrm{Im}\,\lambda$. Dotted horizontal lines: complete graph. The same quantities for RR, BA and WS networks are shown in Fig.~S1 of the Supplementary Material.}
\label{fig:net}
\end{figure}

Fig.~\ref{fig:net} shows simulations on ER networks with $\kbar=5$--$500$ ($N=5000$, $10^5$ steps, 20 runs per point; standard errors are smaller than the symbols in most cases). For $\sigma=10^{-4}$ and $10^{-3}$ the mean activity follows the PA closely over two decades of $\kbar$ (e.g.\ $24.4\pm0.04$ simulated nodes at $\kbar=10$ and $40.4\pm0.06$ at $\kbar=50$ for $\sigma=10^{-4}$), and the amplitude of the oscillations decreases monotonically as $\kbar$ decreases: for $\sigma=10^{-4}$ the standard deviation is $40.1\pm0.4$ nodes at $\kbar=500$, $25.4\pm0.3$ at $\kbar=50$ and $15.5\pm0.1$ at $\kbar=10$. Part of this reduction is due to the lower mean, but the relative amplitude (standard deviation over mean) also decreases, from $0.94$ to $0.64$, in agreement with the stronger damping predicted by the PA. For $\sigma=10^{-5}$, dense networks behave like the complete graph, with large-amplitude oscillations that die out and are reignited (fraction of time with $A=0$ of $0.34$ at $\kbar=500$); at $\kbar=50$ the amplitude is smaller, extinctions are rarer ($0.13$) and the mean activity is higher than on dense networks ($36.5\pm0.2$ against $31.3\pm0.4$ nodes), while at $\kbar\le10$ the activity collapses. Thus, for weak spontaneous activation, a moderate reduction of the connectivity \emph{stabilises} the activity by suppressing the deep troughs of the cycle. In all cases at least $47\%$ of the spectral power lies in the oscillatory band (periods of $250$--$3000$ steps): the oscillation is weakened but not destroyed, because the noise keeps exciting the damped mode. Note that all the networks of Fig.~\ref{fig:net} lie below the PA threshold ($\kc\approx550$ for $\sigma=10^{-5}$, and no sustained cycle for $\sigma>\sigH$), so the oscillations shown there are noise-driven. The amplitude alone decreases with $\kbar$ in both regimes and cannot tell a sustained cycle from quasi-cycles; the size test of Section~\ref{sec:size} shows that below $\kc$ the variance of $A/(wN)$ decreases as $N^{-1}$, while above it saturates.

\begin{figure}[!tbp]
\centering
\includegraphics[width=0.85\textwidth]{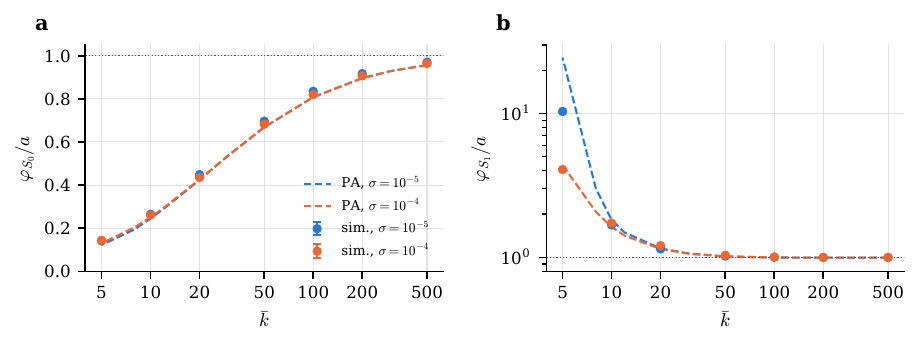}
\caption{Neighbour correlations on ER networks ($N=5000$, $10^5$ steps, 10 runs per point; error bars: standard error, smaller than the symbols) versus the mean degree. (a) Average fraction of active neighbours of naive nodes, $\varphi_{S_0}$, and (b) of informed nodes, $\varphi_{S_1}$, divided by the global active fraction $a=A/N$. Symbols: simulations; dashed lines: pair approximation. Dotted line: well-mixed population.}
\label{fig:corr}
\end{figure}

The damping mechanism of the PA can be checked directly in the simulations. In additional runs (10 per point, sampled every 10 steps) we measured $\varphi_{S_0}$ and $\varphi_{S_1}$, the average fraction of active neighbours of naive and informed nodes, and compared them with the global active fraction $a=A/N$ (Fig.~\ref{fig:corr}); in a well-mixed population both ratios are one. As $\kbar$ decreases, $\varphi_{S_0}/a$ falls to $0.68$ at $\kbar=50$ and to $0.26$ at $\kbar=10$ ($\sigma=10^{-4}$), whereas $\varphi_{S_1}/a$ rises to $1.73$ at $\kbar=10$: naive nodes are depleted around the active ones, which slows down the first stage of the contagion, while informed nodes, which are created by contacts with active nodes, remain next to them. The PA reproduces both ratios within $10\%$ for $\kbar\ge10$ and both values of $\sigma$ (e.g.\ $0.669$ and $0.251$ for $\varphi_{S_0}/a$ at $\kbar=50$ and $10$); larger deviations appear only at $\kbar=5$, where the activity is very low. The PA overestimates the clustering of active nodes among themselves (by a factor of $1.3$ at $\kbar=50$ and $1.6$ at $\kbar=10$ for $\sigma=10^{-4}$, and more for $\sigma=10^{-5}$), which does not enter the contagion rates. The correlations that the PA uses to explain the damping are therefore present in the simulations with the predicted magnitude.

\subsection{A finite-size test of the critical mean degree}
\label{sec:size}

\begin{figure}[!tbp]
\centering
\includegraphics[width=\textwidth]{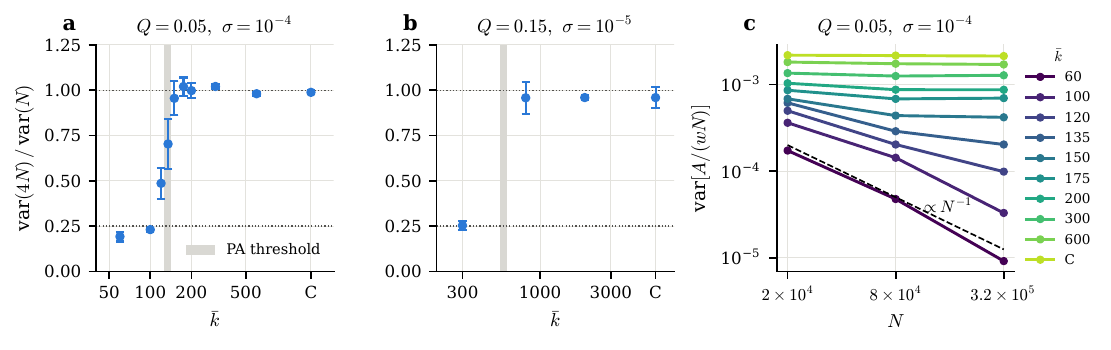}
\caption{Finite-size test of the critical mean degree on ER networks. (a,b) Ratio of the temporal variance of the active fraction $A/(wN)$ between the two largest sizes versus $\kbar$ (C: complete graph): close to $1$ for sustained cycles and to $1/4$ for noise-driven quasi-cycles (dotted lines). Grey band: PA threshold. (a) $Q=0.05$, $\sigma=10^{-4}$, $N=80000$ and $320000$; (b) $Q=0.15$, $\sigma=10^{-5}$, $N=40000$ and $160000$. Error bars from the dispersion between independent graphs. (c) Variance of $A/(wN)$ versus $N$ for case (a); dashed line: $N^{-1}$.}
\label{fig:size}
\end{figure}

The dependence on the population size separates the two regimes. For a sustained cycle the temporal variance of the active fraction $A/(wN)$ tends to a positive constant as $N$ grows, whereas for quasi-cycles around a stable focus it decreases as $N^{-1}$ (Section~\ref{sec:noise}). The ratio $\mathrm{var}(4N)/\mathrm{var}(N)$ is therefore close to one above $\kc$ and close to $1/4$ below it. We carried out this test with an independent C++ implementation of the synchronous rule of Section~\ref{sec:model}, on ER networks with $\nu=0.0016$ kept exactly fixed through $m=\nu wN$, runs of $10^5$ steps (the first $2\times10^4$ discarded) and 2--3 independent graphs per point (Fig.~\ref{fig:size}).

For $Q=0.05$ and $\sigma=10^{-4}$, inside the mean-field oscillatory region and without extinctions, we used $N=20000$, $80000$ and $320000$; the PA predicts $\kc\in(126,141)$, larger than at $\sigma=10^{-6}$ ($65$--$72$, Section~\ref{sec:pahopf}) because spontaneous activation adds damping. The ratio between the two largest sizes is $0.19$ at $\kbar=60$, $0.23$ at $100$, $0.49$ at $120$, $0.70$ at $135$, $0.96$ at $150$, and between $0.98$ and $1.02$ from $\kbar=175$ to $600$ and on the complete graph (Fig.~\ref{fig:size}a): the simulated threshold lies between $120$ and $150$. Fig.~\ref{fig:size}c shows the two behaviours directly: the variance follows $N^{-1}$ for $\kbar\le100$ and is independent of $N$ for $\kbar\ge175$, with a slow crossover close to $\kc$. For the reference case $Q=0.15$, $\sigma=10^{-5}$ ($N=10000$, $40000$, $160000$; PA: $\kc\in(535,598)$) the ratio is $0.25$ at $\kbar=300$ (quasi-cycles) and $0.96$ at $\kbar=800$, $2000$ and on the complete graph (sustained cycles; Fig.~\ref{fig:size}b). The extinctions that dominate small populations in this case (a fraction of time with $A=0$ of $0.10$--$0.19$ at $N=10000$) disappear at $N=160000$ (below $10^{-3}$). The critical mean degree is therefore a simulated result, not only a prediction of the PA.

Near and above the threshold the PA also gives the size of the cycle: for $Q=0.05$, $\sigma=10^{-4}$, the variance of the active fraction along the PA limit cycle is $1.14\times10^{-3}$ at $\kbar=300$ and $1.59\times10^{-3}$ at $\kbar=600$, against $1.28\times10^{-3}$ and $1.70\times10^{-3}$ in the simulations with $N=320000$, and the mean-field cycle gives $2.03\times10^{-3}$ against $2.13\times10^{-3}$ on the complete graph.

\subsection{Contagion structure and turnover on networks}
\label{sec:qnu}

\begin{figure}[!tbp]
\centering
\includegraphics[width=\textwidth]{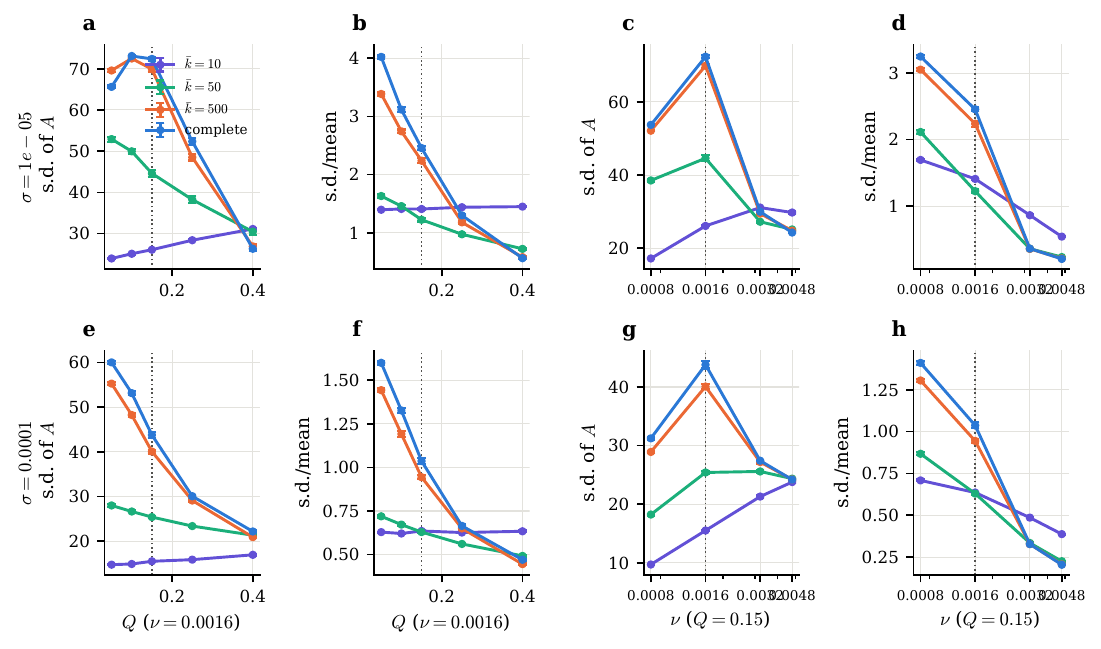}
\caption{Oscillation amplitude on ER networks ($\kbar=10$, $50$, $500$) and on the complete graph versus $Q$ (columns 1--2, $\nu=0.0016$) and $\nu$ (columns 3--4, $Q=0.15$), for $\sigma=10^{-5}$ (top) and $10^{-4}$ (bottom). Columns 1 and 3: standard deviation of the number of active nodes; columns 2 and 4: standard deviation divided by the mean. $N=5000$, 20 runs per point, error bars: standard error. Dotted lines: reference values.}
\label{fig:qnu}
\end{figure}

To see how the two parameters that control the mechanism interact with the connectivity, we simulated ER networks with $\kbar=10$, $50$ and $500$ and the complete graph for $Q=0.05$--$0.40$ (at $\nu=0.0016$) and $\nu=0.0008$--$0.0048$ (at $Q=0.15$), with $\sigma=10^{-5}$ and $10^{-4}$ (Fig.~\ref{fig:qnu}; $N=5000$, 20 runs per point).
\begin{itemize}
\item \textbf{On dense networks $Q$ controls the oscillation, as in the mean field.} On the complete graph the relative amplitude decreases from $4.0$ at $Q=0.05$ to $0.56$ at $Q=0.40$ for $\sigma=10^{-5}$, and from $1.60$ to $0.47$ for $\sigma=10^{-4}$; for $\kbar=500$ the values are almost the same. Smaller $Q$ also means deeper troughs and more frequent extinctions ($53\%$ of the time without activity at $Q=0.05$ against $1\%$ at $Q=0.40$, complete graph, $\sigma=10^{-5}$).
\item \textbf{At low connectivity the effect of $Q$ disappears.} For $\kbar=10$ the relative amplitude is $1.40$--$1.45$ for every $Q$ at $\sigma=10^{-5}$ and $0.62$--$0.64$ at $\sigma=10^{-4}$, and the absolute amplitude even increases slightly with $Q$, because the mean activity does. For $\kbar=50$ the dependence on $Q$ is intermediate. Sparsity therefore erases the signature of the two-stage mechanism: the damping introduced by the correlations dominates over the delay introduced by the informed class. This agrees with the PA, in which no cycle survives at $\kbar=10$ for any $Q$.
\item \textbf{A faster turnover reduces the relative amplitude on every network.} Increasing $\nu$ raises the mean activity ($83$ and $118$ nodes at $\nu=0.0032$ and $0.0048$ on dense networks, against $29$--$42$ at $\nu=0.0016$), shortens the period (from $570$--$650$ steps at $\nu=0.0008$ to $\approx350$ at $\nu=0.0048$ on dense networks) and reduces the relative amplitude to $0.2$--$0.4$ for $\kbar\ge50$ (and to $0.4$--$0.9$ for $\kbar=10$). For $\nu\ge0.0032$ the results for $\kbar=50$, $500$ and the complete graph agree within $10\%$: when the population is renewed quickly the network structure matters little. Conversely, a slow turnover ($\nu=0.0008$) gives the largest relative amplitudes and the most frequent extinctions ($55\%$ of the time without activity on dense networks at $\sigma=10^{-5}$).
\end{itemize}
In summary, sustained, large oscillations need the combination of a predominantly two-stage contagion, a slow turnover and a dense network; reducing any of the three ingredients turns them into small noise-driven fluctuations.

\subsection{Other topologies}
\label{sec:topo}

The Supplementary Material (Section~S1 and Fig.~S1) compares ER networks with RR, BA and WS networks of the same mean degree. ER and RR networks are equivalent within $5\%$ for $\kbar\ge10$, so that the mean degree is the relevant parameter, as assumed by the PA. Degree heterogeneity amplifies the oscillation (the standard deviation is $35\%$ larger on BA than on ER networks at $\kbar=10$, $\sigma=10^{-4}$), whereas clustering damps it ($7.4$ against $15.5$ nodes on WS networks), partly because in clustered networks local neighbourhoods oscillate out of phase with the network as a whole. Measured neighbour correlations (Fig.~S2) show that in WS networks the neighbourhood of active nodes is saturated with informed and refractory nodes, and that in ER and BA networks the depletion of active neighbours around a naive node depends mainly on its own degree, so that the hubs of BA networks see an almost well-mixed neighbourhood. For all topologies the dominant period stays close to the linear period of the mean-field focus.

\section{Bistability and absence of chaos}
\label{sec:other}

For $\sigma=0$ the extinction state is absorbing, and the mean-field equations have a bistable window $\nuT<\nu<\nuF$ ($0.00625<\nu<0.0423$ for the reference parameters), bounded by a transcritical bifurcation and a fold of the endemic branch, in which extinction and an endemic state are both stable; the endemic state loses stability at the Hopf bifurcation $\nuH=0.0017$, below which the oscillations of Section~\ref{sec:mf} are found. Simulations on the complete graph confirm this diagram, and on sparse ER networks the window narrows and moves to larger $\nu$ (Supplementary Material, Section~S3).

As for chaos, the mean-field flow is three-dimensional and the PA has 14 independent variables, so chaos cannot be excluded a priori. A numerical search of the leading Lyapunov exponent \cite{benettin1980} over about $12\,000$ random parameter sets of both descriptions, oscillating ones included (Supplementary Material, Section~S2), found no exponent above the finite-time error of the estimate. This agrees with the simulations of Katriel for the temporary-adoption model with turnover, which revealed no non-stationary dynamics other than a limit cycle \cite{katriel2019}. The irregular time series of the simulations are therefore quasi-cycles generated by demographic noise acting on a regular deterministic skeleton (Section~\ref{sec:noise}), not deterministic chaos.

\section{Discussion and concluding remarks}
\label{sec:disc}

Our main findings can be summarised as follows. (i) In the mean-field description the oscillations of two-stage contagion with turnover are born at a supercritical Hopf bifurcation. Their mechanism is the delay introduced by the informed stage combined with the slow refilling of the naive population: they require a contagion that proceeds mostly through the informed stage ($Q\lesssim0.17$), a slow turnover ($\nu\lesssim\nuH$), an intermediate recovery rate and a large inactive reservoir. Spontaneous activation acts as a perturbation that destroys them above $\sigH$, which is at most $\approx3.7\times10^{-4}$ in the explored region and $4.2\times10^{-5}$ for the reference parameters. Their period increases from the linear value to more than twice that value as $\sigma\to0$, where they become relaxation oscillations. (ii) In finite populations oscillations persist above $\sigH$ as noise-driven quasi-cycles. A linear noise approximation predicts their spectrum and variance quantitatively when the damping is not too weak; close to the bifurcation the fluctuations are nonlinear, smaller than the LNA predicts at moderate $N$, and slower. Below $\sigH$ the deterministic cycle is followed, but its deep troughs cause repeated extinction and reignition in small populations. (iii) Reducing the connectivity introduces correlations that act as an extra damping. Simulations show them directly: at $\kbar=10$ a naive node sees only about a quarter of the active neighbours assumed by the mean field, while informed nodes cluster around active ones, with the magnitude predicted by the pair approximation. In the pair approximation the sustained cycle disappears below a critical mean degree $\kc$, of the order of a few hundred for the reference parameters and of $65$--$72$ for a more two-stage contagion ($Q=0.05$) or a slower turnover ($\nu=0.0008$) at $\sigma=10^{-6}$ (it increases with $\sigma$): the same ingredients that produce the oscillation make it robust to sparsity. A finite-size test with up to $3.2\times10^5$ nodes confirms this threshold: below $\kc$ the variance of the active fraction decreases as $N^{-1}$, above it saturates. In simulations the amplitude of the oscillations decreases monotonically with decreasing $\kbar$, and for weak spontaneous activation a moderate reduction of the connectivity even increases the average activity by suppressing extinctions. (iv) At low connectivity ($\kbar=10$) the amplitude becomes independent of $Q$, i.e.\ sparsity erases the signature of the two-stage mechanism, while a faster turnover reduces the relative amplitude on all networks. As for the topology, at equal mean degree ER and RR networks are equivalent (Supplementary Material); degree heterogeneity amplifies the oscillations and clustering damps them, partly because in clustered networks local neighbourhoods oscillate out of phase with the network as a whole.

In the language of opinion or technology adoption, $\nu$ represents the renewal of the population, $\gamma$ the end of the adoption, $Q$ the probability of adopting at the first exposure and $\sigma$ an exogenous influence such as media or advertising. Within the model, recurrent waves of adoption and abandonment are a feature of novelties that need repeated exposure (small $Q$), in slowly renewed populations (small $\nu$), with little exogenous influence and on densely connected contact networks. A modest exogenous influence turns sustained waves into irregular, noise-driven fluctuations of a similar period; a sparse or clustered contact network damps them further, while a heterogeneous network with hubs amplifies them. These statements are properties of the model, not of any particular data set.

With respect to previous work, for $\sigma=0$ the mean-field system has the structure of the temporary-adoption model of Katriel~\cite{katriel2019}, and the bistable window and the oscillations we find are of the kind reported there. The spontaneous activation, the fraction-of-neighbours rule, the noise-driven oscillations and the network dependence are the new elements. Quasi-cycles sustained by demographic noise are well known in epidemic and predator--prey models \cite{mckane2005,alonso2007}; here they are the generic form of the oscillation once a small exogenous adoption is present. Chang et al.~\cite{chang2026} showed, in a different information--epidemic model, that an external input keeps oscillations alive and that dynamical correlations on static networks suppress them, with a double-edged effect on the mean activity; compared with that work, the two-stage rule, the PA threshold and its finite-size test, the measured neighbour correlations and the role of the topology are new. Pair approximations and edge-based descriptions of network contagion have a long history \cite{keeling1999,eames2002,gleeson2013,kiss2017}, as does the description of transmission on networks in terms of a per-edge transmissibility \cite{newman2002}. For a one-stage process on ER networks, Ref.~\cite{munoz2025} examined the validity of the mass-action description as a function of the connectivity; here a pair approximation captures the mean level, the damping of the oscillation and the neighbour correlations over two decades of $\kbar$.

Our approach has some limitations. The pair approximation treats the network as regular and does not include noise; a linear noise approximation of the pair equations, and a heterogeneous pair approximation for BA networks, would allow a quantitative theory of the quasi-cycles on sparse and heterogeneous networks. The explanation offered for the effect of hubs (Supplementary Material, Section~S1) is supported only indirectly, by the degree dependence of the correlations, and the local-synchrony analysis of the Supplementary Material is based on one run per network. The simulations use 20 runs per point (5 for the two largest complete graphs, 10 for the correlation measurements and 2--3 graphs per point in the finite-size test), and the sweeps in $Q$ and $\nu$ were done on ER networks only. Finally, the synchronous discrete-time update is slightly less damped than its continuous-time limit (its linearised Hopf threshold is $\sigH^{(d)}\approx6.6\times10^{-5}$ instead of $4.2\times10^{-5}$); this is included in the LNA but not in the mean-field and PA bifurcation maps, which refer to the continuous-time equations.

\section*{Acknowledgements}
This is Mario Primicerio's final article. It completes the series on two-stage contagion that he initiated with the Madrid group. He continued to encourage its completion with persistence and sustained interest until the final period of his life. His ideas underpin its conceptual framework. The last developments, computational analyses and revision were completed after his death. All his coauthors acknowledge the example he set as a mathematician and as a man. Luca Meacci acknowledges the support of the Gruppo Nazionale per la Fisica Matematica (GNFM) of the Istituto Nazionale di Alta Matematica (INdAM), Italy, and a fellowship funded by the Italian Ministry and awarded through DINFO, Universit\`a degli Studi di Firenze. This research did not receive any other specific grant from funding agencies in the public, commercial, or not-for-profit sectors.

\section*{CRediT authorship contribution statement}
{\sloppy\textbf{Francisco J.\ Mu\~noz:} Conceptualization, Software, Formal analysis, Writing -- review \& editing. \textbf{Juan Carlos Nu\~no:} Conceptualization, Methodology, Formal analysis, Writing -- original draft, Writing -- review \& editing, Supervision. \textbf{Luca Meacci:} Conceptualization, Software, Formal analysis, Writing -- review \& editing. \textbf{Mario Primicerio:} Conceptualization, Methodology, Formal analysis, Writing -- original draft, Supervision.\par}

\section*{Declaration of competing interest}
The authors declare that they have no known competing financial interests or personal relationships that could have appeared to influence the work reported in this paper.

\section*{Data availability}
The code used to produce the results is provided as supplementary files. The simulation data are available from the corresponding author on reasonable request.

\section*{Declaration of generative AI and AI-assisted technologies in the manuscript preparation process}
During the preparation of this work the authors used Claude (Anthropic) to assist in writing, preparing figures and editing the text. After using this tool, the authors reviewed and edited the content as needed and take full responsibility for the content of the published article.

\appendix

\section{Noise covariance of the linear noise approximation}
\label{app:lna}

For the complete graph, the per-step transitions of the model and their expected numbers per node are: $S_0\to A$ by contact or spontaneously, $(Qa+\sigma)s_0$; $S_0\to S_1$, $(1-Q)as_0$; $S_1\to A$, $as_1$; and $A\to R$, $\gamma a$. Each of them contributes $r_e\,\mathbf{v}_e\mathbf{v}_e^{\mathsf T}$ to $D$, where $r_e$ is the expected number per node and $\mathbf{v}_e$ the change of $(s_0,s_1,a)$, i.e.\ $(-1,0,1)$, $(-1,1,0)$, $(0,-1,1)$ and $(0,0,-1)$. The turnover removes exactly $m=\nu wN$ nodes per step from the non-inactive population, chosen without replacement, and returns $m$ naive nodes; the inflow is deterministic and the outflow is multivariate hypergeometric, which contributes $\nu\left(\mathrm{diag}(\mathbf{x})-\mathbf{x}\mathbf{x}^{\mathsf T}/w\right)$ with $\mathbf{x}=(s_0,s_1,a)$. Correlations between the outcomes of the same node within a step are of higher order in the (small) transition probabilities and have been neglected.

\section{Pair-approximation equations}
\label{app:pa}

Let the events be the transitions $e:\ X_e\to X'_e$ in Table~\ref{tab:events}, with rate $r_e$ per node in state $X_e$. The flux of nodes is $f_e=n_{X_e}r_e$ and the expected number of neighbours in state $Y$ of a node that changes state is $M_{e,Y}$, so that $\delta_{e,Y}=f_eM_{e,Y}/\kbar$. The expected dynamics is
\begin{align}
\dot n_U &=\sum_e f_e\left(-[U=X_e]+[U=X'_e]\right),\\
\dot\pi_{UV} &=\sum_e\Big[\delta_{e,V}\left(-[U=X_e]+[U=X'_e]\right)+\delta_{e,U}\left(-[V=X_e]+[V=X'_e]\right)\Big],
\end{align}
where $[\cdot]$ is $1$ if the condition holds and $0$ otherwise. A transition caused by contact is more likely for nodes with more active neighbours; conditioning on it, the node has on average $1+(\kbar-1)\varphi_X$ active neighbours and $(\kbar-1)q_{Y|X}$ neighbours in any other state $Y$ (contact profile). The initial condition is $\pi_{XY}(0)=n_X(0)\,n_Y(0)$.

\begin{table}[h]
\centering
\caption{Events of the pair approximation. $\varphi_X=\pi_{XA}/n_X$. Contact events: $M_{e,Y}=(\kbar-1)q_{Y|X_e}+\delta_{Y,A}$; free events: $M_{e,Y}=\kbar\,q_{Y|X_e}$.}
\label{tab:events}
\begin{tabular}{lcc}
\toprule
Event & Rate per node $r_e$ & Neighbour profile\\
\midrule
$S_0\to A$ (contagion) & $Q\,\varphi_{S_0}$ & contact\\
$S_0\to S_1$ (contagion) & $(1-Q)\,\varphi_{S_0}$ & contact\\
$S_1\to A$ (contagion) & $\varphi_{S_1}$ & contact\\
$A\to R$ & $\gamma$ & free\\
$S_0\to A$ (spontaneous) & $\sigma$ & free\\
$X\to I$, $X\in\{S_0,S_1,A,R\}$ & $\nu$ & free\\
$I\to S_0$ & $\nu w/(1-w)$ & free\\
\bottomrule
\end{tabular}
\end{table}

\section{Numerical methods}
\label{app:num}

\emph{Mean-field equations} were integrated with LSODA (relative tolerance $10^{-10}$) in the variables $(s_0,s_1,\ln a)$, which allows following trajectories whose activity falls far below one node. Equilibria were obtained by one-dimensional root finding and their stability from the eigenvalues of the Jacobian; the maps of Fig.~\ref{fig:mfmaps} use a grid of $45\times57$ points per plane. The amplitude and period of the limit cycle were measured over the last $1.5\times10^5$ time units of runs of $4\times10^5$ units, started near the equilibrium and from one active node. \emph{Pair-approximation} equilibria were found with Newton's method on the 15 pair variables of the symmetric matrix $\pi_{XY}$ (one equation replaced by the normalisation $\sum\pi_{XY}=1$, which leaves 14 independent variables), continued in $\kbar$ from the mean-field solution, and the Jacobian was obtained by central differences; the zero eigenvalue associated with the normalisation was discarded. \emph{Simulations} use a vectorised Python implementation of the synchronous rule of Section~\ref{sec:model} (NumPy/SciPy, NetworkX for the graphs), equivalent to the original R code; on the complete graph the number of active neighbours of every node is simply $A$ (minus one for active nodes), so no adjacency matrix is needed; spectra were estimated with Welch's method. Neighbour correlations were sampled every 10 steps after the transient, and the ratio $\varphi_X/a$ was computed as $\sum_t\sum_{i\in X}\varphi_i(t)\big/\sum_t n_X(t)\,a(t)$, which equals one in a well-mixed population. Lyapunov exponents were computed as described in Ref.~\cite{benettin1980}, with the variables $(s_0,s_1,\ln a)$ for $\sigma=0$. The code is provided as supplementary files; the finite-size test of Section~\ref{sec:size} used an independent C++ implementation of the same rule. Part of the simulation and analysis code was written with the assistance of Claude (Anthropic) and checked by the authors.

\clearpage
\setcounter{section}{0}\setcounter{figure}{0}\setcounter{table}{0}\setcounter{equation}{0}
\renewcommand{\thesection}{S\arabic{section}}
\renewcommand{\thefigure}{S\arabic{figure}}
\renewcommand{\thetable}{S\arabic{table}}
\renewcommand{\theequation}{S\arabic{equation}}
\section*{Supplementary Material}
This part contains the results on other network topologies (Section~\ref{sec:topo-sm}), the numerical search for chaos (Section~\ref{sec:chaos}) and the bistability in the turnover rate (Section~\ref{sec:bist}). Parameters, rules and observables are those of Section~\ref{sec:model}; the code is provided as supplementary files.

\section{Effect of the topology}
\label{sec:topo-sm}

\begin{figure}[!tbp]
\centering
\includegraphics[width=\textwidth]{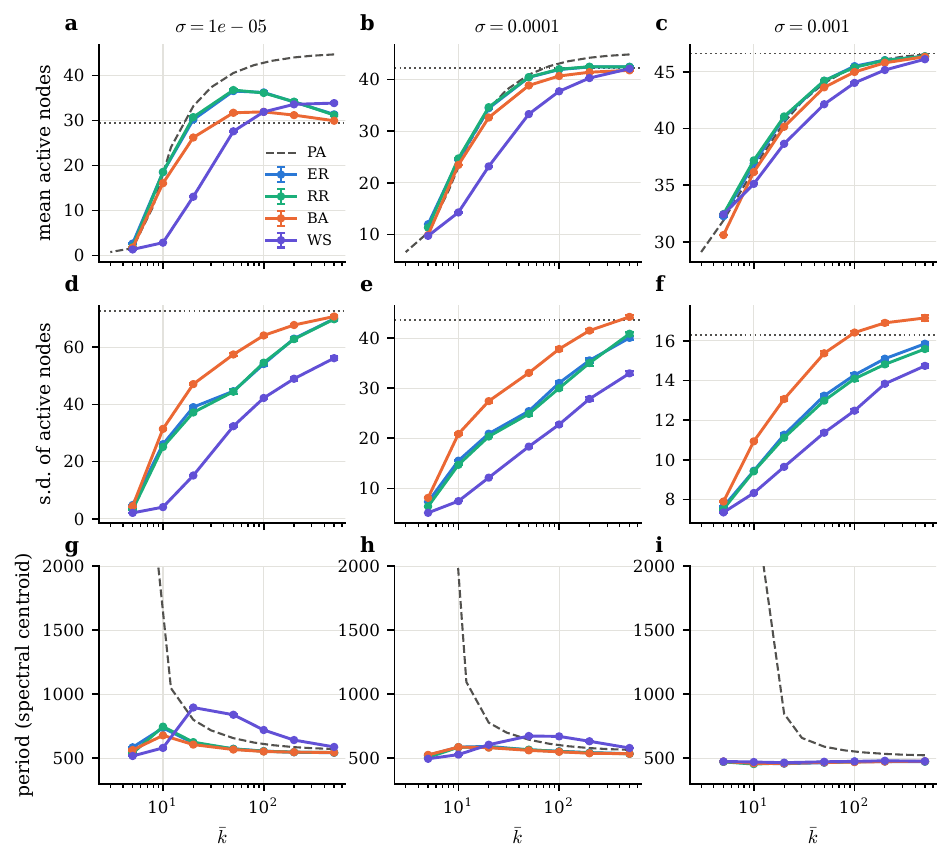}
\caption{Simulations on ER, RR, BA and WS networks ($N=5000$, $10^5$ steps, 20 runs per point; error bars: standard error) versus the mean degree, for $\sigma=10^{-5}$ (left), $10^{-4}$ (centre) and $10^{-3}$ (right). (a--c) Mean number of active nodes; dashed: pair approximation. (d--f) Standard deviation of the number of active nodes. (g--i) Dominant period (centroid of the spectrum in the band of periods $250$--$3000$); dashed: PA linear period $2\pi/\mathrm{Im}\,\lambda$. Dotted horizontal lines: complete graph.}
\label{fig:netall}
\end{figure}

\begin{figure}[!tbp]
\centering
\includegraphics[width=\textwidth]{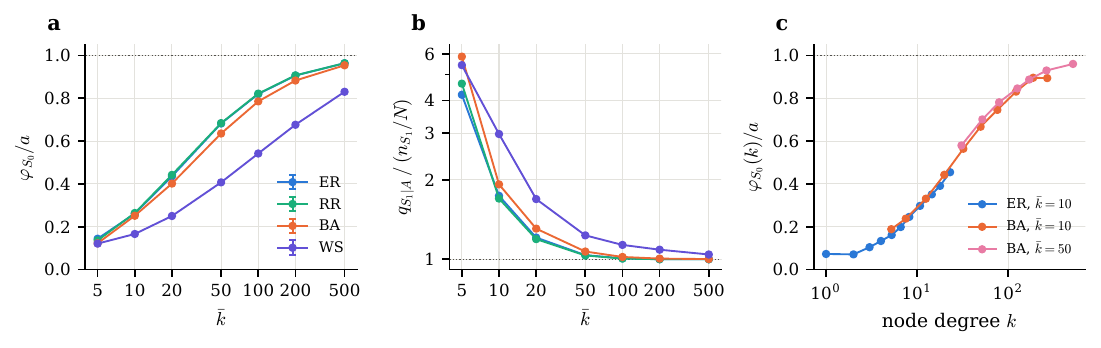}
\caption{Neighbour correlations on ER, RR, BA and WS networks ($N=5000$, $\sigma=10^{-4}$, $10^5$ steps, 10 runs per point). (a) Average fraction of active neighbours of naive nodes divided by the global active fraction $a=A/N$, versus the mean degree. (b) Fraction of informed nodes among the neighbours of active nodes, $q_{S_1|A}$, divided by the fraction $n_{S_1}/N$ of informed nodes in the population. (c) $\varphi_{S_0}/a$ for naive nodes of degree $k$ in ER ($\kbar=10$) and BA ($\kbar=10$, $50$) networks. Dotted lines: well-mixed population.}
\label{fig:corrtopo}
\end{figure}

Fig.~\ref{fig:netall} compares four network ensembles with the same mean degree $\kbar$: Erd\H{o}s--R\'enyi (ER) graphs $G(N,p)$ with $\kbar=p(N-1)$; random regular (RR) graphs with degree equal to the integer nearest to $\kbar$; Barab\'asi--Albert (BA) graphs \cite{barabasi1999} with the integer nearest to $\kbar/2$ edges per new node, so that the mean degree is close to $\kbar$; and Watts--Strogatz (WS) small-world graphs \cite{watts1998}, built from a ring lattice whose degree is the integer nearest to $\kbar$, increased by one if it is odd, with rewiring probability $0.1$. Three features stand out.
\begin{itemize}
\item \textbf{ER and RR networks are almost equivalent}: for $\kbar\ge10$ their mean activity and amplitude differ by at most $5\%$ (e.g.\ $24.4$ and $24.7$ active nodes, standard deviations $15.5$ and $14.7$, at $\kbar=10$, $\sigma=10^{-4}$); larger relative differences appear only at $\kbar=5$, where the activity is very low and ER graphs contain isolated nodes. Their neighbour correlations also coincide (Fig.~\ref{fig:corrtopo}a,b). For these networks the mean degree is the relevant parameter, as assumed by the PA.
\item \textbf{Degree heterogeneity (BA) amplifies the oscillations}: the mean activity is only slightly lower than for ER, but the standard deviation is larger for almost all $\kbar$ and $\sigma$ (the exceptions are $\kbar=5$ and $500$ at $\sigma=10^{-5}$), by $35\%$ at $\kbar=10$ ($20.9\pm0.1$ against $15.5\pm0.1$ nodes, $\sigma=10^{-4}$) and $30\%$ at $\kbar=50$. For $\sigma=10^{-5}$ BA networks also spend more time without activity ($0.38$ against $0.26$ at $\kbar=10$). The average correlations are almost the same as in ER networks ($\varphi_{S_0}/a=0.25$ against $0.26$ at $\kbar=10$, Fig.~\ref{fig:corrtopo}a), so the amplification is not due to weaker correlations on average. They depend, however, on the degree of the node: $\varphi_{S_0}(k)/a$, measured on the naive nodes of degree $k$, falls on the same curve for ER and BA networks and exceeds $0.8$ for $k\gtrsim100$ (Fig.~\ref{fig:corrtopo}c). The hubs therefore see an almost well-mixed neighbourhood. A plausible explanation of the amplification is that hubs, which with the fraction-of-neighbours rule are influenced like any other node but expose many nodes when active, behave as a weakly damped core that synchronises large neighbourhoods; this explanation is consistent with the measured correlations but has not been tested directly.
\item \textbf{Clustering (WS) damps the oscillations and the activity}: at equal $\kbar$ the WS networks have a lower activity and a much smaller amplitude (standard deviation $7.4$ against $15.5$ nodes at $\kbar=10$ and $18.3$ against $25.4$ at $\kbar=50$, $\sigma=10^{-4}$); for $\sigma=10^{-5}$ the activity almost vanishes at $\kbar=10$ ($2.9$ nodes). In a clustered network the neighbours of an active node share many of their neighbours, so that the second exposure tends to fall on nodes that are already informed or refractory: the local saturation of the contagion acts as an additional damping. The measured correlations confirm it (Fig.~\ref{fig:corrtopo}a,b): at $\kbar=10$ a naive node sees $0.17$ of the global active fraction, against $0.26$ in ER networks, and the neighbourhood of active nodes contains $3.0$ times more informed and $2.5$ times more refractory nodes than a random sample of the population, against $1.7$ and $1.5$ in ER networks.
\end{itemize}
The dominant period is close to the linear period of the mean-field focus for all topologies: for $\sigma=10^{-4}$ and $\kbar\ge50$ it is $535$--$565$ steps ($580$--$675$ for WS networks), against $555$ for the focus, and for $\sigma=10^{-3}$ it is $460$--$480$ steps, against $515$; it lengthens moderately at small $\kbar$ for $\sigma\le10^{-4}$. The PA predicts much longer periods at small $\kbar$, but there the oscillatory mode is overdamped and its frequency is not visible in the spectrum.

\paragraph{Local synchrony.} To test whether the network oscillates as a whole or as a collection of partially independent patches, we recorded the state of every node ($N=5000$, $\sigma=10^{-4}$, $4\times10^4$ steps) and computed, for each node, the fraction of active nodes in its closed neighbourhood. After band-pass filtering around the oscillation period (300--1500 steps) we extracted its phase with the Hilbert transform and measured its phase locking to the global signal, $\ell_i=|\langle e^{i(\theta_i-\theta_{\mathrm{glob}})}\rangle_t|$. Because a small neighbourhood is a noisy sample, $\ell_i$ is compared with the same quantity for random sets of nodes of the same sizes, which contain sampling noise but no network structure. The ratio of the mean locking to this null value is $0.92$ and $0.99$ for ER networks with $\kbar=10$ and $50$, $0.90$ for RR and $1.02$ for BA networks with $\kbar=10$: neighbourhoods follow the global oscillation as closely as random samples do, i.e.\ the oscillation is global. For WS networks the ratio is $0.70$ at $\kbar=10$ and $0.69$ at $\kbar=50$: in clustered networks neighbourhoods drift out of phase with the global signal, so that the global oscillation is the average of partially desynchronised local oscillations and its amplitude is reduced. This supports the interpretation of clustering as a source of extra damping.

\section{Search for chaos}
\label{sec:chaos}

The mean-field system is a nonlinear three-dimensional flow, and the PA has 14 independent variables, so chaos cannot be excluded a priori. We estimated the leading Lyapunov exponent \cite{benettin1980} for parameter sets sampled at random ($Q\in[0.01,1]$, $\gamma\in[10^{-3},10^{-1/2}]$, $\nu\in[10^{-4},10^{-3/2}]$, $w\in[0.05,1]$; $\sigma=0$ or $\sigma\in[10^{-9},10^{-2}]$ for the mean field and $\sigma\in[10^{-8},10^{-3}]$ and $\kbar\in\{3,5,10,20,50,150,\infty\}$ for the PA): 5\,000 mean-field sets with $\sigma>0$ (199 oscillating), 2\,800 with $\sigma=0$ (162 oscillating) and 4\,200 PA sets (218 oscillating, all but six with $\kbar\ge50$), plus a regular PA grid at $\kbar\le20$. The largest exponent found was $7.5\times10^{-5}$, of the order of the finite-time error of the estimate, so no chaotic regime was found. The irregular time series of the simulations are quasi-cycles generated by noise acting on a regular deterministic skeleton (Section~\ref{sec:noise}), not deterministic chaos. This is a numerical, non-exhaustive search and cannot prove that chaos is absent.

\section{Bistability in the turnover rate}
\label{sec:bist}

For $\sigma=0$ the extinction state is absorbing. In the mean-field equations, the turnover rate produces a transcritical bifurcation at $\nuT=Qw-\gamma$ ($0.00625$ for the reference parameters) and a fold of the endemic branch at $\nuF=w/[4(1-Q)]-\gamma$ ($0.0423$), so that extinction and an endemic state are both stable for $\nuT<\nu<\nuF$ and the long-time behaviour depends on the initial condition. The endemic state loses stability at a Hopf bifurcation at $\nuH=0.0017$, below which the oscillatory regime of Section~\ref{sec:mf} is found. Simulations on the complete graph ($N=5000$, $\sigma=0$) confirm the diagram: started with a large seed ($A_0=250$) the activity survives for $0.003\le\nu\le0.03$ (the upper limit lies below $\nuF$, as expected for a finite population close to the fold) at a level that follows the stable mean-field branch (e.g.\ $203$ nodes at $\nu=0.01$, mean field $202$), whereas a single seed reaches the endemic state only with probability $0.1$--$0.2$. For $\nu\le0.002$ the activity always dies out: close to and below $\nuH$ the approach to the endemic state is oscillatory and its first trough falls below one node; the extinction threshold decreases with $N$ towards $\nuH$. On sparse ER networks ($\kbar=30$ and $8$) the window of persistent activity narrows and moves to larger $\nu$, but a single seed becomes more efficient.

\end{document}